\documentclass[superscriptaddress,print,longbibliography]{revtex4-2}

\usepackage{graphicx}
\usepackage{dcolumn}
\usepackage{bm}
\usepackage{multirow}
\usepackage{color}
\usepackage{lineno}
\usepackage[utf8]{inputenc}
\usepackage{upgreek}
\usepackage{amsmath}
\newcommand{\mr}[1]{\mathrm{#1}}  

\begin{document}

\title{Reversal of nanomagnets by propagating magnons in ferrimagnetic yttrium iron garnet enabling nonvolatile magnon memory}

\author{Korbinian Baumgaertl}
\affiliation{Laboratory of Nanoscale Magnetic Materials and Magnonics, Institute of Materials (IMX), \'Ecole Polytechnique F\'ed\'erale de Lausanne (EPFL), 1015 Lausanne, Switzerland}
\author{Dirk Grundler}\email{dirk.grundler@epfl.ch}
\affiliation{Laboratory of Nanoscale Magnetic Materials and Magnonics, Institute of Materials (IMX), \'Ecole Polytechnique F\'ed\'erale de Lausanne (EPFL), 1015 Lausanne, Switzerland}
\affiliation{Institute of Electrical and Micro Engineering (IEM), \'Ecole Polytechnique F\'ed\'erale de Lausanne (EPFL), 1015 Lausanne, Switzerland}

\date{\today}

\begin{abstract}
\textbf{Despite the unprecedented downscaling of CMOS integrated circuits, memory-intensive machine learning and artificial intelligence applications are limited by data conversion between memory and processor. There is a challenging quest for novel approaches to overcome this so-called von Neumann bottleneck. Magnons are the quanta of spin waves and transport angular momenta through magnets. They enable power-efficient computation without charge flow and would solve the conversion problem if spin wave amplitudes could be stored directly in a magnetic memory cell. Here, we report the reversal of ferromagnetic nanostripes by spin waves which propagate through an underlying spin-wave bus made from yttrium iron garnet. Thereby, the charge-free angular momentum flow is stored after transmission over a macroscopic distance. We show that spin waves can reverse large arrays of ferromagnetic stripes at a strikingly small power level of nW. Combined with the already existing wave logic, our discovery is path-breaking for the new era of magnonics-based in-memory computation and beyond von Neumann computer architectures.}
\end{abstract}


\maketitle

Charge-based electronics exploiting semiconductors and metals have led to the digital age and given rise to ever increasing data-processing power during the last decades. The significant development slows down however due to overheating of processors at increased clock frequencies as well as the data-transfer bottleneck between charge-based processors and storage devices \cite{Manipat2018,Sebastian2020}. So far, storage devices exploit reprogrammable states in materials classes which differ from the semiconductors and therefore require signal conversion. Waves for data processing like photons in optical computers already avoid, both, charge flow and the excessive Joule heating at high clock frequencies. However, even for optical computers the currently implemented von-Neumann device architecture still provokes the signal-conversion bottleneck, particularly in case of memory-intensive machine learning applications. Magnonics which is based on charge-free angular momentum flow via spin waves (magnons) promises to overcome the dilemmas. Making use of the bosonic quasiparticles  \cite{BoseEin2006}, it offers low-power consuming wave-based computing  with microwave signals up to THz frequencies \cite{Khitun_2010,ChumakMagTrans,Mahmoud2020}. A technology platform for disruptive information technologies is foreseen if the so far elusive nanomagnet switching by propagating spin waves would be experimentally demonstrated. This switching process in magnonic circuits would be path-breaking for both non-volatile storage of wave signals without conversion losses and in-memory computation \cite{IRDS2020,Islam_2019,Sebastian2020}. Experiments presented in Ref. \cite{Wang2019b} are encouraging in that spin waves in an antiferromagnet excited by incoherent spin torques reversed an integrated micromagnet. However, such spin waves can not be used for wave logic. In Ref. \cite{Han1121}, spin waves coherently excited in a magnon conduit by a microwave antenna induced the modification of magnetic domains. Here, the required power was however in the mW regime. We follow a different approach in that we excite spin waves in the ferrimagnetic insulator yttrium iron garnet ${\textrm{Y}}_{3}{\textrm{Fe}}_{5}{\textrm{O}}_{12}$ (YIG) by radiofrequency (RF) signals and observe the magnon-induced reversal of ferromagnetic nanostripes integrated on its top surface (Fig. \ref{Fig1}a). The shape anisotropy of the long and narrow permalloy (Py) stripes (green) allows for bistable magnetic states. Thereby, they represent magnetic bits with states "0" and "1", and enable the nonvolatile storage of information (Fig.\,\ref{Fig1}b, I and II). The underlying YIG (blue) allows us to transmit coherent magnons over macroscopic distances as their decay lengths in thin YIG are on the few 10 to 100 $\mu$m length scale \cite{Yu2014SR,Maendl2017}. They are far larger than electron scattering lengths in semiconductors and metals. We note that samples similar to the one shown in Fig. \ref{Fig1}a have thoroughly been investigated recently in view of the so-called grating coupler effect. Different groups explored nanostripe arrays on YIG in their saturated states for the emission and detection of short-wave magnons  \cite{Yu2016,Liu2018,PhysRevB.100.104427,KBaumgaertl2020}. The magnon modes were transmitted over long distances \cite{Liu2018} and imaged in real space by means of x-ray based magnetic microscopy \cite{KBaumgaertl2020}. However, the groups did not investigate if states of ferromagnetic nanostripes were different before and after spin-wave excitation at different power levels. Here, we report experiments that we performed on YIG spin-wave buses (delay lines) with integrated grating couplers (Fig. \ref{Fig1}a) and led to the discovery of irreversible switching of the nanostripes by propagating magnons. By combining power-dependent spin-wave spectroscopy and magnetic force microscopy we evidence the storage of magnon signals in the YIG-based magnonic circuit containing ferromagnetic stripes. Importantly, the magnon-induced reversal reported here is induced by magnons which are transported over a macroscopic distance ($> 20~\mu$m). This goes beyond Refs. \cite{Wang2019b,Slon2010,Suresh2021,Guo2021,Zheng2022} which considered distances on the nm length scale. Our finding is key for implementing data storage in wavelogic or neuromorphic computing architectures in magnonics \cite{Papp2021}. Spin waves in YIG have already been shown to read out different magnetic states of integrated ferromagnetic stripes (Fig.\,\ref{Fig1}b, III) \cite{Liu2018,Qin2021}. Combined with our discovery, in-memory computing and a complete read-and-write process of non-volatile magnetic bits via magnons in YIG circuits are now possible.

\subsection*{Reversal of ferromagnetic stripes on YIG observed by power-dependent spin-wave spectroscopy}
\begin{figure*}[htbp!]
	\includegraphics[width=1\textwidth]{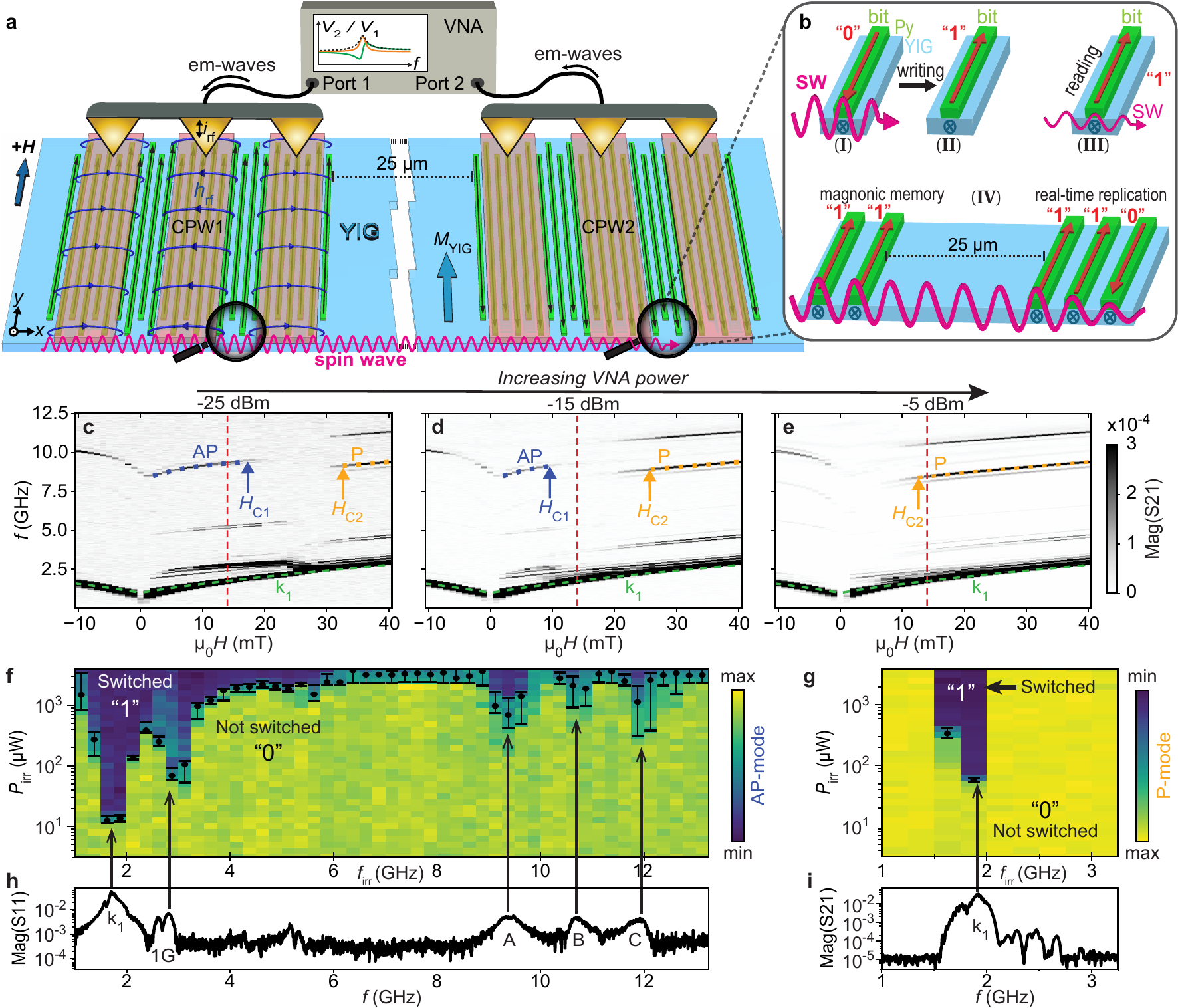}%
	\caption{\textbf{Magnonic memory effect\textendash Reading and writing of magnetic bits by spin waves.} \textbf{a} Two lattices of bistable magnetic bits (Py nanostripes) underneath CPWs on an insulating YIG film. \textbf{b} Depending on the SW amplitude bit writing (I and II), reading (III) and data replication (IV) are achieved without charge flow. \textbf{c-e} Transmission signals Mag(S21) taken at three different power levels $P_\mathrm{irr}$. Analyzing signal strengths of mode branches (blue and orange dashed lines), fields $H_{C1}$ and $H_{C2}$ are extracted reflecting bit reversal. The red dashed lines indicate $+14~\mr{mT}$. Yield of bit writing (dark) underneath \textbf{f} CPW1 and \textbf{g} CPW2 by propagating SWs at $\mu_{0}H_\mathrm{B}=+14~$mT. The 50 \% transition power levels $P_\mr{C1}$ and $P_\mr{C2}$, respectively, are marked as black dots. The error bars indicate the 70 \% and 30 \% transitions (Methods). \textbf{h} Mag(S11) at $P_{\rm sens}$ and \textbf{i} Mag(S21) taken at +14 mT. In \textbf{i}, bits at CPW1 and CPW2 were magnetized to state "1".}\label{Fig1}
\end{figure*}

We studied different YIG-based magnonic circuits with integrated ferromagnetic nanostripes which showed magnon-induced reversal of nanomagnets reported here. We explored ferromagnetic nanostripes with widths between 50 and 200 nm, lengths up to 27 $\mu$m, edge-to-edge separations down to 50 nm and different orientations with respect to spin-wave emitters. In this publication we focus on the magnonic circuit shown in Fig.\,\ref{Fig1}a consisting of two nanostripe arrays which form grating couplers \cite{Yu2013} on YIG underneath coplanar waveguides (CPWs). The insulating YIG with magnetization $M_{\rm YIG}$ is 100 nm thick. The emitter (left) and detector (right) CPWs are connected to a vector network analyzer (VNA) which provides electromagnetic (em) waves at GHz frequencies at port 1. In such magnonic circuits information can be encoded in either the SW amplitudes or phases \cite{Kostylev2005,Schneider2008, Khitun_2010,Csaba2017}.
The magnetic component of the em wave, $\mathbf{h}_{\rm rf}$ (field lines sketched in Fig. \ref{Fig1}a), is converted into propagating spin-wave (SW) signals at the CPW via the torque $-\mid\gamma\mid\mu_0\mathbf{M}_{\rm YIG}\times\mathbf{h}_{\rm rf}$, where $\gamma$ is the gyromagnetic ratio and $\mu_0$ the vacuum permeability. We read out the SW signals both in reflection configuration at the emitter CPW1 and, after transmission through the YIG circuit, at the detector CPW2. The measured scattering parameters S11 and S21, respectively, are phase-resolved voltage signals induced by spin-precessional motion. The VNA offers a frequency resolution of 1 Hz and precisely adjusted power levels. In Fig.\,\ref{Fig1}a, 25-$\upmu$m- and 27-$\upmu$m-long ferromagnetic nanostripes are integrated between CPWs and YIG (see Methods and Supplementary Fig. \ref{MagMemSup1}). These are bistable nanomagnets made from the ferromagnetic metal $\mathrm{Ni_{81}Fe_{19}}$ (permalloy, Py). In earlier works such bistable nanomagnets were investigated in the linear regime and gave rise to the grating coupler (GC) effect \cite{Yu2013,Yu2016,Liu2018,KBaumgaertl2020}. Here, we report a nonlinear phenomenon in that GC stripes are reversed by spin waves which propagate through YIG. We show that thereby the storage of SW signals occur. To highlight the additional functionality of the ferromagnetic nanostripes we consider them to represent magnetic bits with either state "0" (magnetization vector $\mathbf{M}_{\rm Py}$ pointing along $-y-$direction) or state "1" ($\mathbf{M}_{\rm Py}$ along $+y-$direction). Our findings hence demonstrate that propagating SWs in YIG can not only sense ("read") the state of such magnetic bits (Fig. \ref{Fig1}\,b) as reported before \cite{Liu2018,Qin2021}, but also reverse ("write") magnetic states at appropriate power levels which can be still in the linear SW excitation regime.
The 100-nm-wide stripes considered here are arranged in separated periodic arrays with a common lattice constant $a=200~$nm. For the read-out (sensing) of "0" and "1" states we apply a small power level $P_{\rm sens}$ of only $-25~$dBm at port 1 (or smaller) and make use of the GC effect in S11 and S21 spectra. It has already been shown that the signal strength and precise frequency of GC modes depended decisively on the field direction and relative magnetization directions of nanostripes and YIG \cite{Liu2018,PhysRevB.100.104427}. When the configuration of emitter and detector grating couplers were not symmetric the detection of emitted GC modes was not possible \cite{Maendl2018}. Using an in-plane magnetic field $H$ we reset the magnetic states of stripes to a well-defined initial state ("erase cycle") before performing subsequent magnon-induced reversal experiments.\\
First, we show that the nanostripes in the magnonic circuit are switched from "0" to "1" by SWs exhibiting a certain amplitude. Figure \ref{Fig1}\,c-e displays the magnitude Mag(S21) of the frequency-dependent voltages at CPW2 as a function of field when measured for three different VNA power levels $P_\mathrm{irr}$ at CPW1. Dark branches indicate SW propagation. Initially, all nanostripes were magnetized along the $-y-$direction ("erased") by applying $\mu_{0}H=-90~\mr{mT}$. Then $\mu_{0}H$ was increased in 1~mT steps to the $+y-$direction. For all powers $P_\mathrm{irr}$, Mag(S21) contained a prominent and continuous low-frequency branch (marked by green dashed lines) attributed to the $k_1$-mode of SWs excited by CPW1 directly in the YIG film \cite{KBaumgaertl2020}. The $k_1$-branch changed slope close to $\mu_{0}H\simeq 0~\mr{mT}$, indicating that the magnetization $M_{\rm YIG}$ of the soft magnetic YIG film was aligned with the external magnetic field for values of +1~mT and above. High frequency branches between roughly 7.5 and 12.5 GHz are attributed to the GC effect. They showed distinct transitions (jumps) in resonance frequencies. These frequency variations are known to indicate {\em irreversible switching} of Py nanostripes \cite{Liu2018,PhysRevB.100.104427}. Following Refs. \cite{Nozaki2007,Topp2009}, the jumps occur at the switching fields at which the magnetization direction of nanostripes reverse. We focus on the two prominent branches marked by blue and orange dashed lines, which, in the following, we denote antiparallel (AP) and parallel (P) mode, respectively (Supplementary Methods). $H_{C1}$ ($H_{C2}$) defined the critical field value for which the signal strength of the AP-mode (P-mode) decreased (increased) to 50~\% of its maximum value (see Methods and Supplementary Fig. \ref{Fig2}). From Fig.~\ref{Fig1}\,c to e, $H_{C1}$ and $H_{C2}$ diminished significantly with increasing $P_\mathrm{irr}$ indicating a nonlinear process and the reversal of nanostripes. Before presenting the magnon-induced reversal by means of frequency-resolved switching-yield maps introduced below, it is instructive to present magnetic force microscopy (MFM) data. We perform MFM at remanence to demonstrate that the magnon-induced reversal of ferromagnetic nanostripes is non-volatile.

\subsection*{Evidence of non-volatile state reversal after magnon transport in YIG}
\begin{figure*}[tbh!]
	\includegraphics[width=\textwidth]{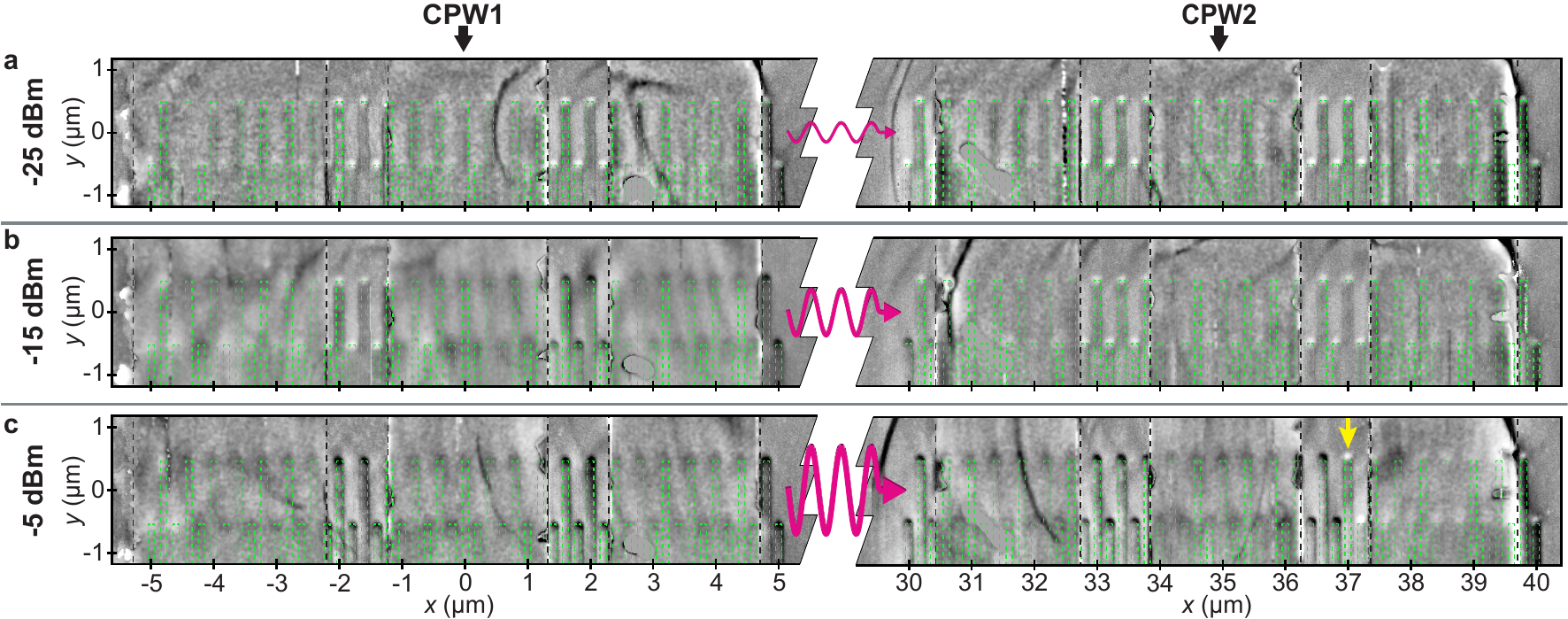}%
	\caption{\textbf{Non-volatile storage of spin-wave signals over macroscopic distance.} MFM measurements taken after the sample with erased bits (state "0", shown as white end) was irradiated at $\upmu_0 H_\mr{B}=+14~\mr{mT}$ with different powers \textbf{a} $P_\mr{irr}=-25$~dBm, \textbf{b} $-15$~dBm, and \textbf{c} $-5$~dBm in the frequency range 0.1~GHz to 12.5~GHz. In this range, spin waves are excited in YIG. Reversed magnetic bits (highlighted by broken lines) appear with a black end. In \textbf{c}, beyond $x=37~\upmu\mr{m}$ (marked by a yellow arrow) stripes remained magnetized in state "0" (white end), consistent with the decay of the SW amplitude when propagating away from the emitter CPW1.}
\label{Fig3}
\end{figure*}

Figure \ref{Fig3}\,a displays an MFM image taken in the remnant configuration after we performed a broadband VNA measurement with $P_\mr{irr}=-25$~dBm at $\mu_{0}H_\mathrm{B}=+14~$mT on the sample with initially "reset" magnetic stripes.
The MFM data show nanostripes at CPW1 and CPW2 which are magnetized still along $-y-$direction and opposite to $\mathbf{M}_{\rm YIG}$, consistent with our definition of the AP branch (cf. Fig. \ref{Fig1}\,c). Note that $\mu_{0}H_\mathrm{B}$ was intentionally below the evaluated onset field of quasistatistic reversal of magnetic stripes (Supplementary Fig. \ref{MagMemSup2}) in order to reveal SW-induced reversal processes. Figure \ref{Fig3}\,b shows the MFM data after applying an increased power level of $P_\mr{irr}=-15$~dBm (31.6~$\upmu\mr{W}$) at $\mu_{0}H_\mathrm{B}=+14~$mT (cf. Fig. \ref{Fig1}\,d). The remnant MFM image shows nanostripes which on the right side of the signal line of CPW1 are reversed to the $+y-$direction (state "1"). On the left side with $x<-1.1~\upmu\mr{m}$ they are still magnetized along $-y-$direction. Fallarino \emph{et al.} \cite{6466529} showed that SWs were excited nonreciprocally by a CPW for the same field geometry as used in our experiments. A large (small) SW amplitude was reported for the right (left) side of the signal line with $x>0$ ($x<0$). We hence attribute the spatially inhomogeneous switching of nanostripes to the non-uniform (nonreciprocal) amplitude distribution of SWs excited in YIG. Stripes at CPW2 were still magnetized in $-y-$direction for $P_\mr{irr}=-15$~dBm, with the exception of the two stripes on the left edge. For this experiment with an intermediate power level, a large number of stripes of the two gratings was magnetized in opposite directions. This magnetic disorder explained the vanishingly small signal amplitude for GC modes in Fig. \ref{Fig1}\,d at 14 mT \cite{Maendl2018}. After a VNA measurement on again reset magnetic stripes with a high power of $P_\mr{irr}=-5$~dBm (c.f. Fig. \ref{Fig1}\,e) the MFM data show that the majority of stripes below the emitter CPW1 was switched (Fig. \ref{Fig3}\,c). Below CPW2, about two thirds of the magnetic stripes were switched to "1" as well. Only the stripes farthest away from CPW1 showed still the initial state "0" (yellow arrow). The spatial variation agrees with a magnon-induced reversal process which requires a threshold amplitude of the SWs propagating under CPW2. As the amplitude of the propagating SWs decayed in positive $x$-direction, the required amplitude was not available for a large propagation distance from CPW1. The MFM-detected magnetic states hence represent a non-volatile memory which records whether a certain threshold SW amplitude has locally been present (state "1") or not (state "0"). Thereby, wave-based computational results can be stored without signal conversion. The reestablished magnetic order underneath both CPWs by high VNA power in Fig. \ref{Fig3}\,c explains the observation of the branch P at +14 mT in Fig. \ref{Fig1}\,e. We note that the significant asymmetry of switched nanostripes observed in Fig. \ref{Fig3}\,b and the subsequently presented experimental data allow us to exclude microwave-induced heating as the trigger for reversal. Heating in our continuous wave experiments would be most likely mirror-symmetric with respect to $x=0$ of the emitter CPW. The direct electromagnetic cross talk between CPW 1 and CPW2 is -50 dB ($10^{-5}$), i.e., vanishingly small.

\subsection*{Power-dependent reversal field reduction}

\begin{figure}[htb!]
	\centering\includegraphics[width=86mm]{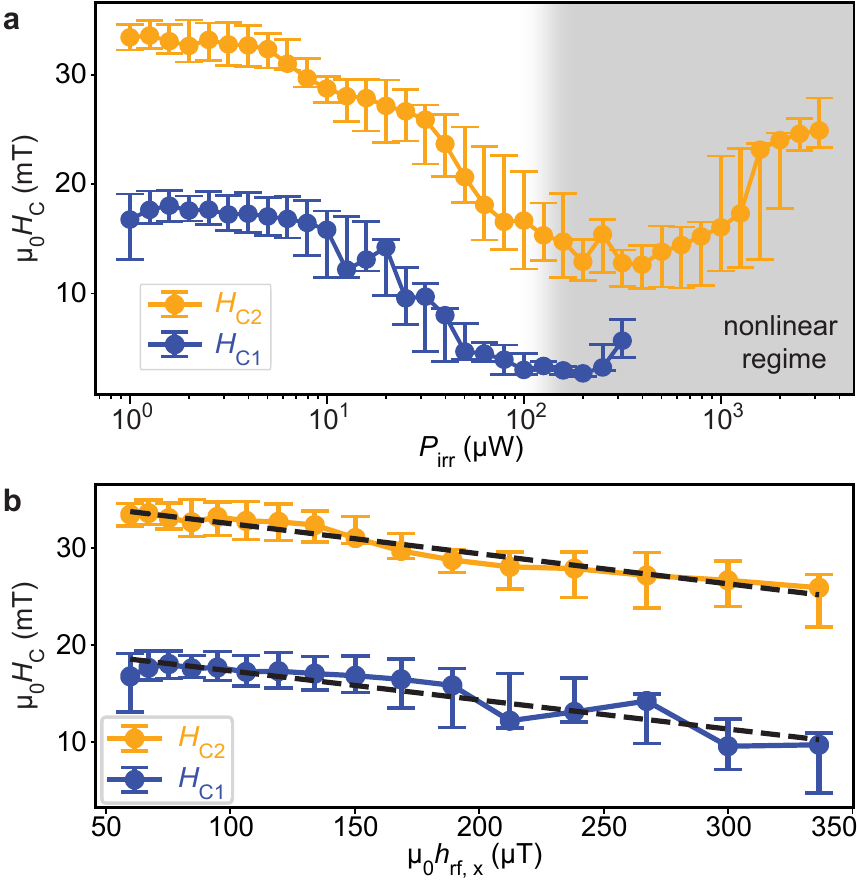}
	\caption{\textbf{Efficiency analysis of reversal field reduction by linear and nonlinear spin waves.}
\textbf{a} Dependence of critical fields $H_{C1}$ (solid blue line) and $H_{C2}$ (solid orange line) extracted from VNA spectra on $P_\mr{irr}$ applied in a broad frequency range from 0.1~GHz to 12.5~GHz. The power region attributed to nonlinear effects in the SW modes is shaded in light gray. \textbf{b} $H_{C1}$ and $H_{C2}$ as a function of the evaluated in-plane dynamic field amplitude $\upmu_0 h_\mathrm{rf, x}$ in the linear SW regime. The error bars represent the 30 \% and 70 \% switching field values. The straight lines are guides to the eye.
	}\label{Fig4}	
\end{figure}

Figure~\ref{Fig4}\,a summarizes how critical fields $H_{C1}$ (blue line with dots) and $H_{C2}$ (orange line with dots) extracted from VNA data depended on the power used in VNA spectroscopy. We find that $H_{C1}$ ($H_{C2}$) decreases from 16.8~mT (33.4~mT) at $P_\mathrm{irr}=1~\mu\mathrm{W}$ to 2.7~mT (12.9~mT) at $200~\mu\mathrm{W}$. This corresponds to a reduction of $H_{C1}$ by 84 \% and $H_{C2}$ by 61 \%.
Surprisingly, for large $P_\mathrm{irr}$ we find again an increase of $H_{C1}$ and $H_{C2}$. We attribute this observation to the onset of nonlinear SW scattering (Supplementary Fig. \ref{MagMemSup3}), which limits the peak amplitude of SWs emitted from CPW1 \cite{PhysRevB.86.054414}. For $P_\mathrm{irr}$ above -9~dBm (shaded with gray in Fig. \ref{Fig4}\,a) we found a red-shift of the $k_{1}$-mode resonance frequency (Supplementary Fig. \ref{MagMemSup3}), which is an indication for nonlinear SW excitation \cite{doi:10.1063/1.2756481}. The reoccurring increase of critical fields is counterintuitive if heating of Py nanostripes by high-power microwaves would be at the origin of the observed reversal phenomenon. If we plot $H_{C1}$ and $H_{C2}$ versus $\upmu_0 h_\mathrm{rf,x}=\upmu_0\sqrt{P_\mathrm{irr.}/(2Z_0 w_\mathrm{L}^2)}$ \cite{PhysRevLett.99.207202} (Fig.~\ref{Fig4}\,b), the slopes (straight lines) show that a small in-plane RF field amplitude of only 0.35 mT provokes a critical field reduction by about 10 mT ($Z_0=50~\Omega$ and $w_\mathrm{L}=2.1~\mu$m, see Supplementary figures). Hence, the RF-field-excited SWs in YIG reduce the stripes' switching fields about $30$ times more efficiently than a static Oersted field applied oppositely to $\mathbf{M}_{\rm Py}$.
\\

\subsection*{Magnon-induced reversal of magnetic nanostripes at low power level}

To evaluate quantitatively the critical powers $P_\mr{C}$ needed to reverse ferromagnetic nanostripes ("switching-yield map"), we excited SWs at well-defined frequencies $f_\mathrm{irr}$, varied the excitation power $P_\mathrm{irr}$ and explored after each power step whether the magnetic configuration of nanostripes changed (Supplementary Fig. \ref{Fig2}). We defined $P_\mr{C1}$ ($P_\mr{C2}$) as the critical power value at which mode branches exhibited 50~\% of their maximum signal strength, indicating that a considerable number of Py nanostripes under CPW1 (CPW2) were switched; the results for an applied field $\mu_{0}H_\mathrm{B}=+14$~mT are summarized by black dots in Fig.~\ref{Fig1}\,f (g) (datasets for a separate sample are shown in Supplementary Fig. \ref{Sample2}). We considered further $H_\mathrm{B}<H_\mathrm{C1}(-25~$dBm$)=16.8~$mT and obtained consistent diagrams (not shown). We find the minimum $P_\mr{C1}$ of 12.6~$\upmu$W when applying an RF signal to CPW1 in the frequency window $f_\mathrm{irr}=1.5$~GHz to 1.75~GHz. This window includes the peak frequency of the $k_1$-mode (1.70~GHz) with a wavelength of 7.4~$\mu$m (Methods), as evidenced by the SW absorption spectrum Mag(S11) (Fig.~\ref{Fig1}\,h). We note that this wavelength is much longer than both the width of stripes and the period $a$ of arrays. The wavelength is not commensurable with the stripe period $a$. The second lowest $P_\mr{C1}$ of 68.9~$\upmu$W occurs for $f_\mathrm{irr}=2.75$~GHz to 3.0~GHz. This frequency range coincides with the frequency of the GC mode named $k_1+1G$. This mode exhibits a wavelength of 195~nm which is a value close to period $a$ (for mode identification see Supplementary  Fig. \ref{Simul} and Ref. \cite{KBaumgaertl2020} for the P-mode). The microwave power $P_\mr{prec}$ which is transferred into magnetization precession is evaluated by $P_\mr{prec}(f_\mathrm{irr})=P_\mr{irr}(f_\mathrm{irr})\cdot[\mr{Mag(S11)}(f=f_\mathrm{irr})]^2$. We obtain critical power values $P_\mr{C1,prec}=36~\mr{nW}$ and $P_\mr{C1,prec}=3.4~\mr{nW}$ for magnetic stripe reversal by means of the $k_1$ mode and the short-wave magnon labelled $k_1+1G$, respectively. The switching-yield map features three further dips at high frequency marked as A, B, and C. They occur at relatively large power values $P_\mr{C1}$ of 0.69~mW, 1.91~mW and 1.13~mW for $f_\mathrm{irr}$ near 9.25~GHz, 10.75~GHz and 11.75~GHz, respectively. These dips agree with eigenresonances of the Py stripes. At such frequencies, the concomitant reversal of Py stripes underneath CPW1 is attributed to the conventional microwave-assisted magnetization reversal (MAMR). MAMR of individual nanomagnets was pioneered in 2003 \cite{Thirion2003} and then applied to mesoscopic Py magnets in e.g. Refs. \cite{Nozaki2007,Topp2009,Nembach2007,PhysRevLett.99.207202}. These micromagnets were not exposed to a propagating spin wave. Instead, they were irradiated directly by an RF signal. Its frequency hit exactly the eigenfrequency of the Py. We measured the direct electromagnetic crosstalk between the CPWs. At CPW2, it amounted to -50 dB. In other words, the directly irradiated microwave power was five (!) orders of magnitude lower and too small for MAMR at CPW2. The dips in Fig. \ref{Fig1}f that occur near 8 GHz will not be discussed in the following. The important features are the ones at low frequency in Fig. \ref{Fig1}f and g. They go beyond MAMR in that (1) we excite spin precession in YIG and not in the Py nanostripes, (2) different spin-wave frequencies in the YIG induce switching of Py nanostripes and (3), nanostripes more than 25 micrometres away from the emitter CPW reverse (Fig. \ref{Fig1}g).\\
$P_\mr{C2}$ (Fig. \ref{Fig1}g) is the critical power for magnetic stripe reversal under CPW2. We find $P_\mr{C2}=58.4~\upmu$W when exciting the $k_1$-mode in YIG (Fig.~\ref{Fig1}\,i). Its eigenfrequency was slightly blue-shifted compared to spectra taken for $P_\mathrm{irr}<P_\mr{C1}$ evidencing the concomitant magnetic stripe reversal by the $k_1$-mode at CPW1. Hence, a SW mode near 2 GHz can "write" states "1" at CPW1 and perform a real-time replication of these bit-states "1" about 25~$\mu$m away [Fig.~\ref{Fig1}(b), panel IV]. Similar to Ref. \cite{Qin2021} we observed that reversed stripes changed the transmitted SW amplitude. Such a signal variation was recently functionalized in a neuronal network based on propagating magnons \cite{Papp2021}. The maximum distance over which reversal is possible is determined by the intermediate damping as will be discussed in the following.\\
The minimum $P_\mr{C2}$ at CPW2 is about 4.5 times larger than the minimum $P_\mr{C1}$. We attribute this discrepancy to the damping of the propagating $k_1$ SW mode. Assuming $I(s)=I(x=0)\exp{(-s/l_{\rm d})}$ and a decay length $l_{\rm d}$ of $23~\upmu\mr{m}$, the intensity $I$ of a SW would decay by 1/4.5 over a distance $x=s=35~\upmu\mr{m}$, i.e., the center-to-center distance between CPW1 and CPW2. Based on the intrinsic Gilbert damping $\alpha_{\rm i}=9\times 10^{-4}$ of the YIG film, we calculate a decay length of $99~\upmu\mr{m}$ for SWs with a wave vector $k_1$ in unpatterned YIG (Supplementary Fig. \ref{LineWith}). It is reasonable to assume that the effective damping parameter in the magnonic circuit of Fig.~\ref{Fig1} was larger due to additional magnon scattering and radiative losses \cite{Schoen2016}. We hence consider $l_{\rm d}=23~\mu$m to be a reasonable value. The mode $k_1+1G$ has a smaller velocity than mode $k_1$ (Supplementary Fig. \ref{LineWith}b) and thereby a further decreased decay length. This consideration explains why the maximum available power of the VNA was not sufficient for nanostripe reversal via mode $k_1+1G$ underneath CPW2. \\
\subsection*{Discussion}
The magnon-induced reversal of bistable ferromagnetic nanoelements via different propagating spin wave modes such as $k_1$ and $k_1+1G$ in YIG is a key asset when aiming at in-memory computation and beyond von-Neumann device architectures based on multi-frequency magnonic circuits. Further studies are needed to understand the microscopic mechanism behind the magnon-induced reversal reported here and thereby predict the theoretical limit of minimum power consumption. But already at this stage several considerations can be made. CPWs and grating couplers emit spin waves at specifically designed wave vectors and due to the spin wave dispersion relation in YIG a specific frequency band is thereby realized. Importantly we observe low-power magnon-induced reversal for both a pure CPW mode (mode $k_1$) and a GC mode ($k_1+1G$). The nature of the mode does not seem to play a decisive role. The observed frequency selectivity is so far attributed to the microwave-to-magnon transducer used and not the reversal process itself. A decade ago, J. C. Slonczewski pointed out that magnons in a ferrimagnet/ferromagnet hybrid structure have the potential to provide spin-transfer torque for nanomagnet reversal more efficiently than charge flow \cite{Slon2010}. His theoretical predictions reflected that for a given amount of energy more bosonic magnons than fermionic electrons could be generated. If transferred to an adjacent nanomagnet, each of these magnons provided an angular momentum of $\hbar$ for reversal. Beyond this so-called magnon torque, spin precessional motion in YIG was found to inject a spin-polarized current $\mathbf{j}_{\rm s} \propto \mathbf{M}_{\rm YIG}\times d\mathbf{M}_{\rm YIG}/dt$ into Py \cite{Hyde2014}. The spin current $\mathbf{j}_{\rm s}$ displayed a nonreciprocal behavior \cite{Iguchi2013} and its amplitude increased with the SW wave vector in YIG \cite{Iguchi2013,SandwegSaitoh2010}. The spatial symmetry of our MFM data of reversed stripes is consistent with both the predicted magnon torque or $\mathbf{j}_{\rm s}$. Furthermore, the $k_1$ and $k_1+1G$ modes have indeed shown a characteristically different power level for reversal. One might hence speculate that magnon torque \cite{Han1121,Wang2019b,Guo2021,Zheng2022} and/or spin current \cite{Suresh2021,Iguchi2013} could play a role for the switching of Py stripes. We note however that we observe similar reversal also in hybrid structures with a 5 nm thick intermediate layer between YIG and Py consisting of either the insulator SiO$_2$ or metal Cu \cite{Mucchietto2022}. These further experiments and results will be presented elsewhere, but render the so far discussed scenarios unlikely as their efficiency should vary significantly with the different intermediate layers. We speculate that the dipolar magnetic fields of propagating spin waves introduce torques on Py magnetic moments. As the frequencies of SW modes $k_1$ and $k1+1G$ in YIG are well below the eigenfrequencies of the Py nanostripes (Fig. \ref{Fig1}f and g and Supplementary Fig. \ref{Sample2}), they do not fulfill the condition for resonant chiral absorption discussed theoretically in Ref. \cite{Kruglyak2021}. A possible scenario might instead consist of SW-induced nucleation of a domain wall followed by its propagation due to the applied static magnetic field. We now compare the power levels reported here with the pioneering spin-transfer torque experiment in Ref. \cite{Albert2000} performed in zero and non-zero applied fields $H$. In Ref. \cite{Albert2000}, a directly injected electrical (el) current was used to switch {\em one single} Co nanomagnet of a small volume $(60\times130\times2.5)~{\rm nm^3}$. A minimum power $P^*_{\rm el}\approx 6~\mu{\rm W}$ was required. In our study, we evaluated a critical power of $P_\mr{C1,prec}=3.4~\mr{nW}$ when reversing {\em a few 10} Py nanostripes of a much larger volume $(100\times26000\times20)~{\rm nm}^3$ {\em each}. In our so-far non-optimized magnonic devices, the required power levels are also orders of magnitude below the ones recently reported for domain-wall manipulation by SWs in Py stripes and the torques produced by incoherent magnons in a thin antiferromagnetic barrier \cite{Han1121,Wang2019b}. Stimulated by Ref.~\cite{Slon2010}, we argue that the bosonic nature of magnons is in general advantageous to initiate magnet reversal. In our experiments, the magnons in YIG had frequencies $f$ around 2 GHz. They were hence of low energy $hf$ where $h$ is the Planck constant. Consequently, a small power can generate a large number of identical low-energy magnons in a short time which then produce torques for reversal via different scenarios outlined above. We expect further reduced critical power levels for magnetic bit reversal and smaller $P_\mr{C,prec}$ by optimizing microwave-to-magnon transducers, modifying the shape or volume of nanomagnets and engineering the interface between magnetic bit and magnon conduit via e.g. an antiferromagnetic layer.



\section*{Methods}
\subsection*{Sample Fabrication}
The samples were prepared on a 100 nm thick yttrium iron garnet (YIG) film, which was grown by liquid phase epitaxy and purchased from the Matesy GmbH in Jena, Germany. On top of the insulating YIG, a 20 nm thick metallic Py ($\mr{Ni_{81}Fe_{19}}$) film was deposited via electron beam evaporation. The grating pattern was defined via electron beam lithography (EBL) using negative hydrogen silsesquioxane resist and transferred into the Py film by $\mr{Ar^+}$ ion beam etching. The etching time was optimized to ensure minimal overetching into the YIG film. Coplanar waveguides were prepared by EBL lift-off processing using a PMMA/MMA double layer positive resist and evaporation of Ti/Cu (5 nm/110 nm).

\subsection*{Spin wave spectroscopy with a vector network analyzer}
SW spectroscopy measurements were conducted with a VNA (Keysight PNA N5222A) in a microwave probe station with integrated in-plane magnetic field control. The magnetic field was created by electromagnets and stabilized via a Hall-sensor controlled feedback loop. High-frequency coaxial cables with $50~\mr{\Omega}$ impedance and microwave probes were used to connect port 1 and port 2 of the VNA to CPW1 and CPW2, respectively. For all conducted measurements, microwave power $P_\mr{irr}$ (or $P_\mr{sens}$) was only applied to port 1, i.e. only the scattering parameters S11 and S21 were measured. The front panel jumpers of the PNA N5222A were adjusted at port 2 to provide increased forward (S21) dynamic range. For measurement of the microwave spectra at different power levels (cf. Fig. \ref{Fig1} \textbf{c-e} and raw data for Fig. \ref{Fig4}) the frequency was scanned from 10~MHz to 12.5~GHz in 3.1~MHz steps. An IF bandwidth of 1~kHz was used. For the switching-yield measurements (cf. Fig. \ref{Fig1} \textbf{f} and \textbf{g}) $f_\mr{irr}$ was scanned in 2.5~MHz steps. An IF bandwidth of 1~kHz and a dwell time of 50~ms was used. The magnitudes of the reflection and transmission scattering parameters Mag(S11) and Mag(S21) were calculated as the square‐root of the sum of the respective squared real and squared imaginary parts, after removing nonmagnetic background signals, which did not depend on the applied external magnetic field. $\mr{Mag(S11)}^2$ is the measure of the relative absorbed power and used to calculate the power $P_\mr{prec}=P_\mr{irr}\cdot\mr{Mag(S11)}^2$ transferred from the CPW into the spin-precessional (prec) motion in the magnonic circuit. The most prominent SW excitation of CPW1 occurred at a wave vector $k_1=0.85~{\rm rad}~\mu$m$^{-1}$ \cite{KBaumgaertl2020}. The wavelength of the corresponding $k_1$ mode in YIG amounted to about 7.4~$\mu$m. The periodic array of Py nanostripes gave rise to a reciprocal lattice vector $1G$ and enabled the excitation of short-wave magnons by the grating coupler effect \cite{Yu2013}. Its wave vector amounted to $k_{1G}=2\pi/a=2\pi/(0.2~\mu$m)~$=31.4~{\rm rad}~\mu$m$^{-1}$. The absorption spectrum Mag(S11) in Fig.~\ref{Fig1}\,h showed a double peak structure for the resonant feature labelled with 1G. We attributed the smaller peak on the low-frequency side of 1G to a grating coupler mode with $-k_1+1G$ and the larger high-frequency peak to $k_1+1G$. The mode $k_1+1G$ exhibited an absolute wave vector of $k_{k1+1G}=k_1+k_{1G}$. Its wavelength amounted to $2\pi/k_{k1+1G}=195$~nm under the stripe array \cite{KBaumgaertl2020}. This value was close to $a$ due to $k_1\ll k_{1G}$. The GC mode $-k_1+1G$ corresponded to a wavelength of 206 nm under CPW1. Considering its larger absorption strength we assumed mode $k_1+1G$ to induce mainly the reversal of nanostripes in the text.

\subsection*{Analysis of Switching Fields}
To determine the power and frequency dependence of the discovered switching process, the following measurement routine was conducted: The stripes were first magnetized downwards with $\mu_{0}H =-90~\mathrm{mT}$ and then the field was increased slowly to a positive bias field of $\mu_{0}H_\mathrm{B} =14~\mathrm{mT}$. This field value was below $\mu_0H_{\rm C1}$ at -25 dBm, i.e., below the evaluated minimum field of the switching field distribution (Supplementary Fig. \ref{MagMemSup2}). We monitored Mag(S21) with a low sensing power $P_\mathrm{sens}=-25$~dBm. This power level was chosen such that $P_\mathrm{sens}$ itself was insufficient to induce switching. While staying at $\mu_{0}H_\mathrm{B}=14~$mT, we applied $P_\mathrm{irr}$ in a small frequency window $f_\mathrm{irr}$ with a range of 0.25~GHz. $P_\mathrm{irr}$ was increased from -30~dBm to 6~dBm in  1~dBm steps. After each power step, Mag(S21) was measured with $P_\mathrm{sens}$ in a frequency window from 7.5~GHz to 12.5~GHz. Thereby we covered the branches indicating the magnetic configuration of the stripes. After each such sequence the sample was reset to the initial state (bits were erased) for the next frequency $f_{\rm irr}$. Supplementary Fig.~\ref{Fig2}a shows exemplarily Mag(S21) measured as a function of $P_\mathrm{irr}$ with $1.5~\mr{GHz}\leq f_\mathrm{irr}\leq 1.75~\mr{GHz}$. With increasing $P_\mathrm{irr}$ the AP-mode vanishes and the P-mode appears. We evaluate mode strengths as a function of $P_\mathrm{irr}$ (Supplementary Fig. \ref{Fig2}b) by integrating Mag(S21) in the vicinity of the AP-mode (enclosed by blue dashed lines) and P-mode (enclosed by orange dashed lines). We define $P_\mr{C1}$ ($P_\mr{C2}$) as the critical power value for which the signal strength of the AP-mode (P-mode) is at 50 \% of its maximum value (Supplementary Fig. \ref{Fig2}b), indicating that 50~\% of the Py nanostripes in the grating under CPW1 (CPW2) were switched. Further we extract the values for which 30 \% and 70 \% of the signal strength are reached as a measure of the switching power distribution (error bar). The measurements were repeated for different $f_\mathrm{irr}$ altered from 1~GHz to 13.25~GHz in 0.25~GHz wide windows. We quantified the relative numbers of reversed nanostripes by integrating the signal strengths in Mag(S21) spectra in the vicinity of the AP- and the P-mode (Methods). At the low power level of $P_{\rm sens}=-25~$dBm we analyzed the quasistatic switching field (SF) distribution of Py nanostripes under CPW1 (Supplementary Fig. \ref{MagMemSup2}a and b) and found a full width half maximum $\mr{SF_{S11,f}}-\mr{SF_{S11,i}}$ of roughly 13 mT centered around a switching field value of $(\mr{SF_{S11,f}}+\mr{SF_{S11,i}})/2=21.5~$mT (Supplementary Fig. \ref{MagMemSup2}c). With increasing $P_{\rm irr}$ we observed a decrease in SF values and a narrower distribution (Supplementary Fig. \ref{MagMemSup2}c) consistent with the writing of bits reported in the manuscript.

\section*{Acknowledgments}
The authors acknowledge discussions with Andrea Mucchietto, Shreyas Joglekar, Michal Mruczkiewicz, and Anna Fontcuberta i Morral.

\subsection*{Funding}
The research was supported by the SNSF via grant number 163016.

\subsection*{Author contributions}
D.G. and K.B. planned the experiments and designed the samples. K.B. prepared the samples and performed the experiments. K.B. and D.G. analyzed and interpreted the data. K.B. and D.G. wrote the manuscript.

\subsection*{Competing interests}
The authors declare that they have no competing interests.

\subsection*{Data availability}
The datasets generated and/or analysed during the current study are available from the corresponding author on reasonable request.

\section*{Additional information}
\textbf{Supplementary Information} is available for this paper.
\\
\textbf{Correspondence} and requests for materials should be addressed to D.G. (email: dirk.grundler@epfl.ch).\\
Reprints and permissions information is available at www.nature.com/reprints.\\
\newpage


\newpage
\renewcommand{\figurename}{\textbf{Supplementary Fig.}}
\setcounter{figure}{0}
\subsection*{SUPPLEMENTARY INFORMATION}
\begin{figure*}[htb!]
	\includegraphics[width=0.80\textwidth]{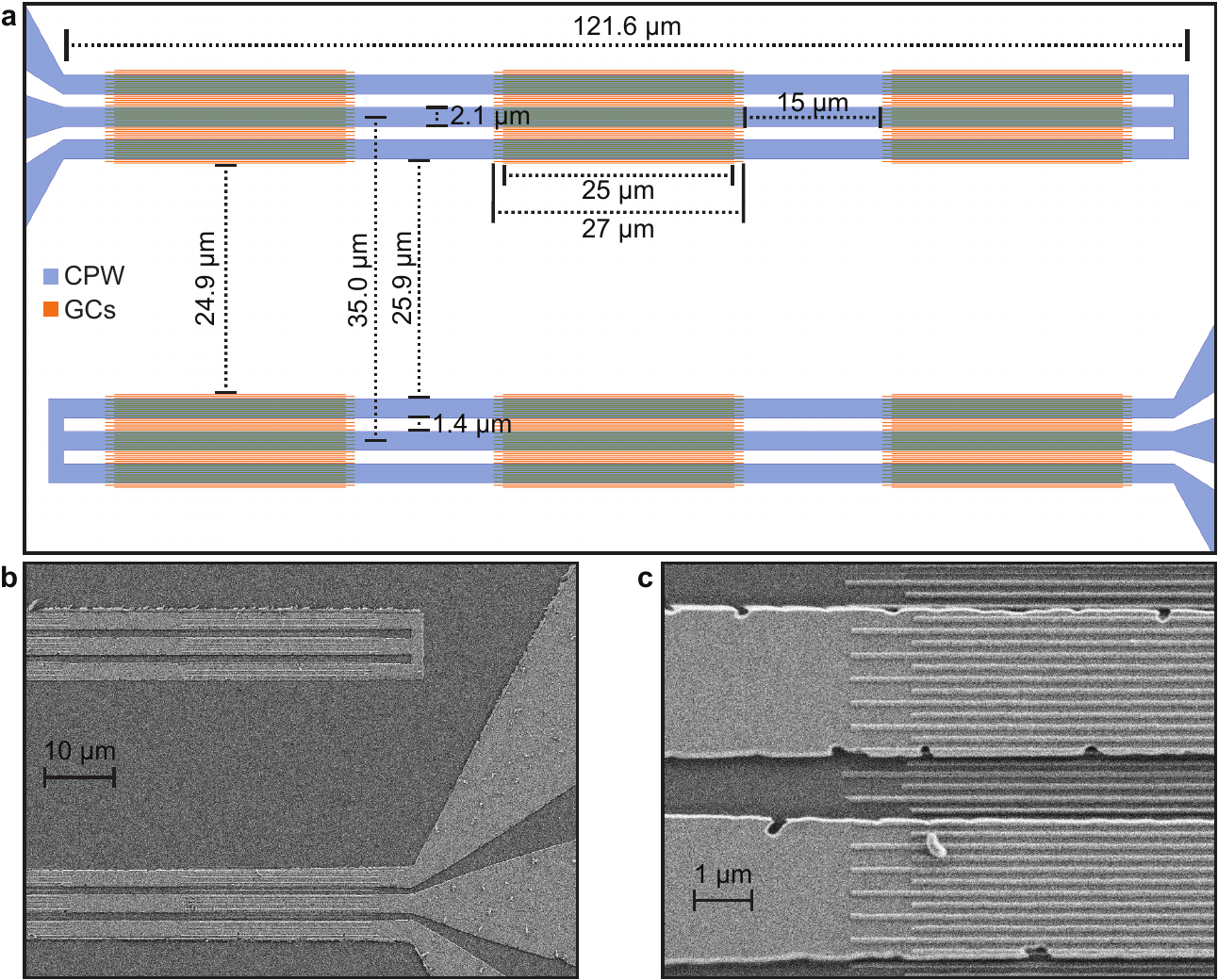}%
	\caption{\textbf{$\mid$~Layout and scanning electron microscopy images of the sample.} \textbf{a} Sample  layout. Two CPWs (blue color) with a total length of $121.6~\upmu\mathrm{m}$ are separated by a center-to-center distance of $35~\upmu\mathrm{m}$. Below each CPW three $25~\upmu\mathrm{m}$ long Py grating couplers (GCs) are arranged. Every second stripe of the GCs is prolonged on both sides by $1~\upmu\mathrm{m}$ and then its length amounted to 27 $\mu$m. The edge-to-edge separation of neighboring GCs underneath one and the same CPW amounts to $15~\upmu\mathrm{m}$ to avoid dipolar interaction. \textbf{b} Scanning electron microscopy image showing both CPWs with GCs. \textbf{c} Zoom in on the end of one of the GCs.}\label{MagMemSup1}
\end{figure*}

\begin{figure*}[tb!]
	\includegraphics[width=70mm]{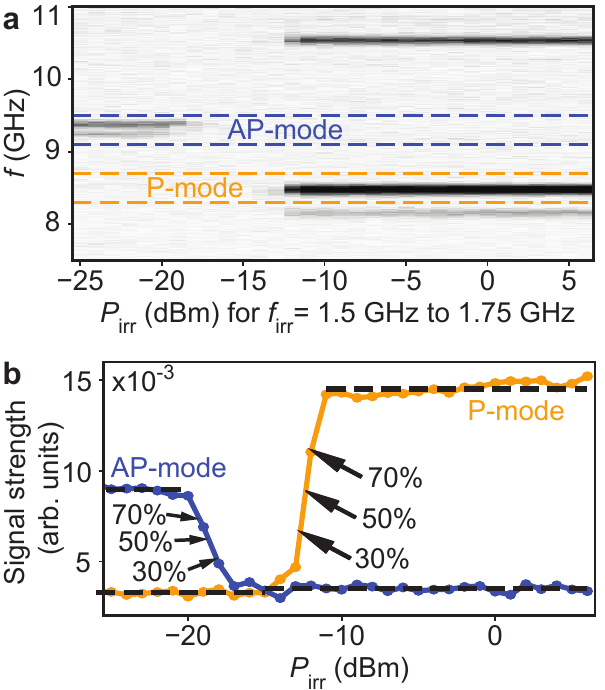}%
	\caption{\textbf{$\mid$~Evaluation of critical power levels $P_\mr{C1}$ and $P_\mr{C2}$.}
		 \textbf{a} Stripes were magnetized along $-y-$direction and a small power of $P_\mathrm{sens}=-25$~dBm was used to measure Mag(S21). Staying at $\mu_{0}H_\mathrm{B} =+14~\mathrm{mT}$ a microwave signal of power $P_\mathrm{irr}$ was applied in 0.25~GHz wide windows exemplary plotted for $f_\mathrm{irr}=1.5$~GHz to 1.75~GHz and increased in a stepwise manner. After each increment of $P_\mathrm{irr}$, Mag(S21) was measured with the small power $P_\mathrm{sens}$ between 7.5~GHz to 12.5~GHz. A transition from the AP-mode (enclosed by blue dashed lines) to the P-mode (enclosed by orange dashed lines) as function of $P_\mathrm{irr}$ was observed. \textbf{b} We quantify critical power levels $P_\mr{C1}$ and $P_\mr{C2}$ by evaluating the 50 \% signal strength transitions of AP- (blue) and P-mode (orange), respectively. We argue that the decay of SW amplitudes below CPW1 and CPW2 and the inhomogeneous stripe reversal in the grating couplers observed by MFM explain why AP- and P-mode transitions appear gradual and are not sharp step functions.
	}\label{Fig2}
\end{figure*}

\begin{figure*}[htb!]
	\includegraphics[width=\textwidth]{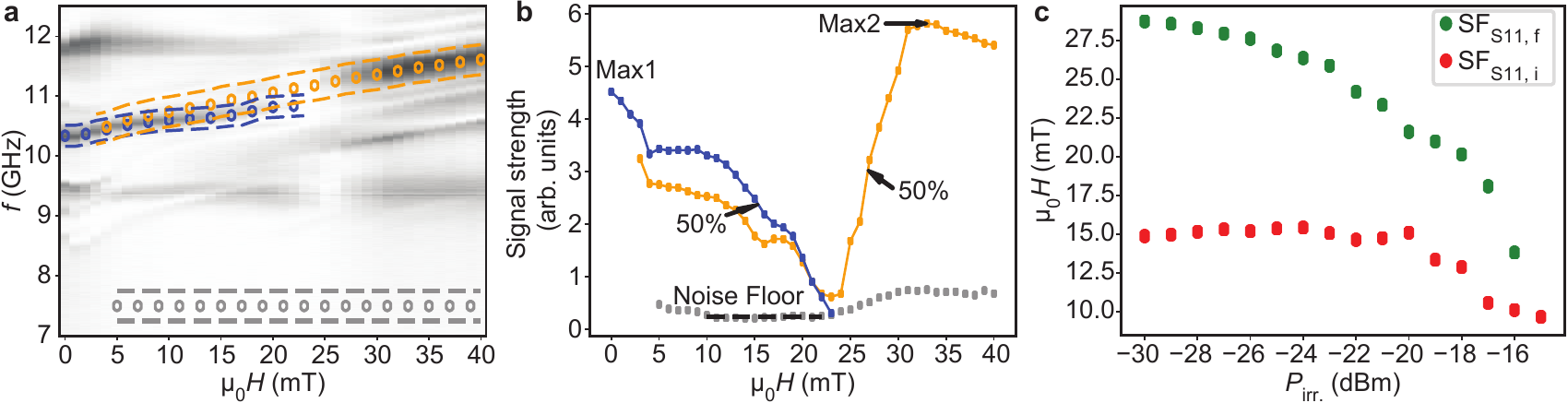}%
	\centering
	\caption{\textbf{$\mid$~Power-dependent switching field distribution of Py nanostripes underneath CPW1.} \textbf{a} Exemplary Mag(S11) for $P_\mathrm{irr}=-25$~dBm. We track a characteristic mode of the AP-configuration (blue dots) and the P-configuration (orange dots). \textbf{b} We quantify the switching field $\mr{SF_{S11,i}}$ ($\mr{SF_{S11,f}}$) by evaluating the decay (rise) of the signal strength to 50~\% of its maximum value with respect to the noise floor (estimated from the region between the gray dashed lines in \textbf{a}). \textbf{c} $\mr{SF_{S11,f}}$ diminished more significantly than $\mr{SF_{S11,i}}$ with $P_\mathrm{irr}$. Above $P_\mathrm{irr}\geq -15$~dBm, $\mr{SF_{S11,i}}$ and $\mr{SF_{S11,f}}$ were separated by less than 1~mT and automatic evaluation with the previously introduced methodology (cf. \textbf{a, b}) was not performed.
}\label{MagMemSup2}
\end{figure*}

\begin{figure*}[htb!]
	\includegraphics[width=0.6\textwidth]{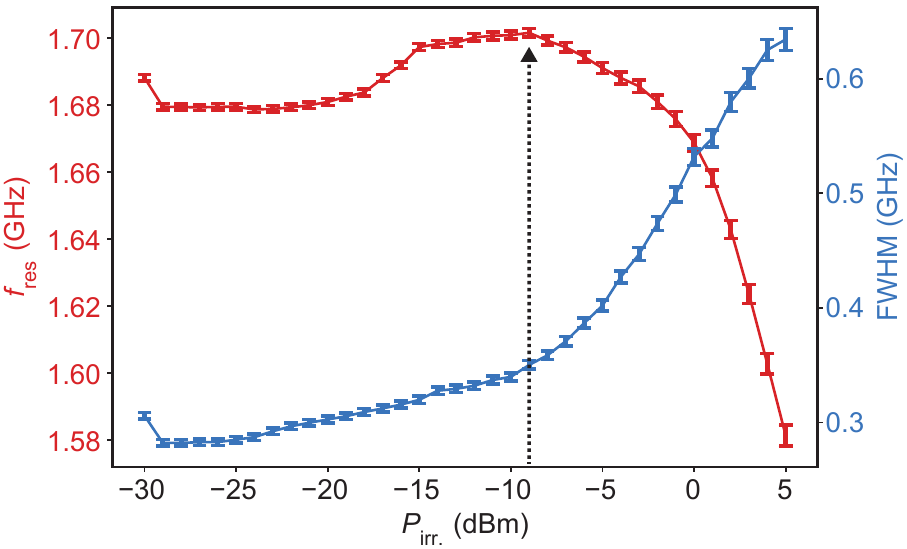}%
	\centering
	\caption{\textbf{$\mid$~Onset power level of nonlinear spin-wave regime.} Power dependence of resonance frequency $f_\mathrm{res}$ (red color) and full width half maximum (FWHM) of the $k_1$ mode resonance in Mag(S11) extracted by fitting with a Lorentzian function at $\upmu_0 H = -10$~mT. Above $P_\mathrm{irr}=-9$~dBm (indicated by dashed arrow) $f_\mathrm{res}$ starts to decrease significantly with increasing power. Around the same $P_\mathrm{irr}$ the FWHM starts to increase significantly. These two observations are attributed to the onset of the nonlinear regime of SW excitation. }\label{MagMemSup3}
\end{figure*}

\begin{figure*}[htb!]
	\includegraphics[width=1\textwidth]{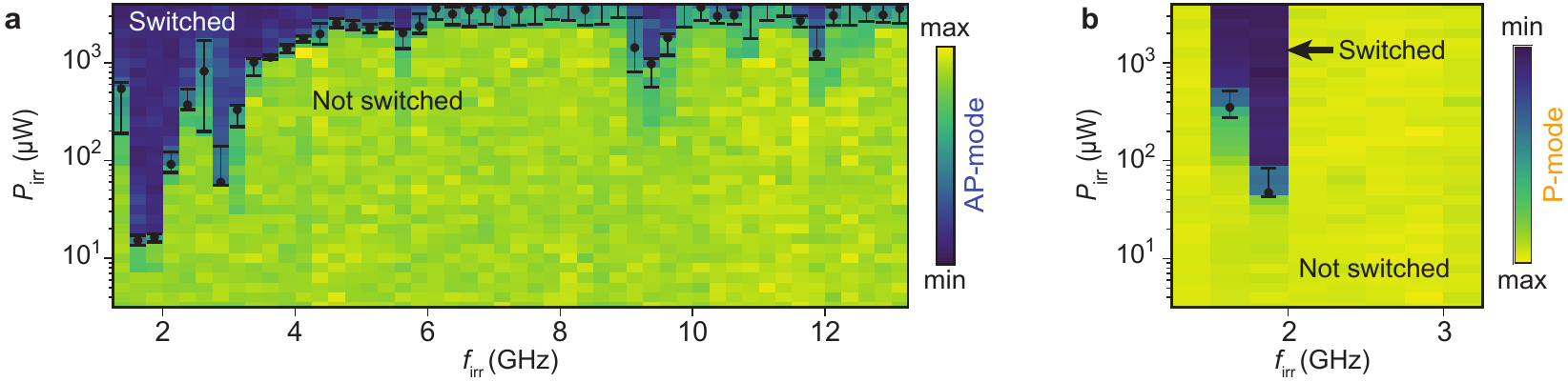}%
	\centering
	\caption{\textbf{$\mid$~Switching-yield maps at $\mu_{0}H_\mathrm{B}=14~\mr{mT}$ for an independently prepared second sample with nominally identical Py nanostripe arrays.} The YIG film was taken from the same wafer, but, evaporation of the Py film and subsequent process steps were conducted separately from the sample presented in the main text. The yields for magnetic bit writing of this second sample out of a different fabrication batch agree well with the findings presented in Fig. 1 of the main text. The agreement is true for writing magnetic bits, both, under \textbf{a} CPW1 and \textbf{b} CPW2.}\label{Sample2}
\end{figure*}

\begin{figure*}[htb!]
	\includegraphics[width=0.82\textwidth]{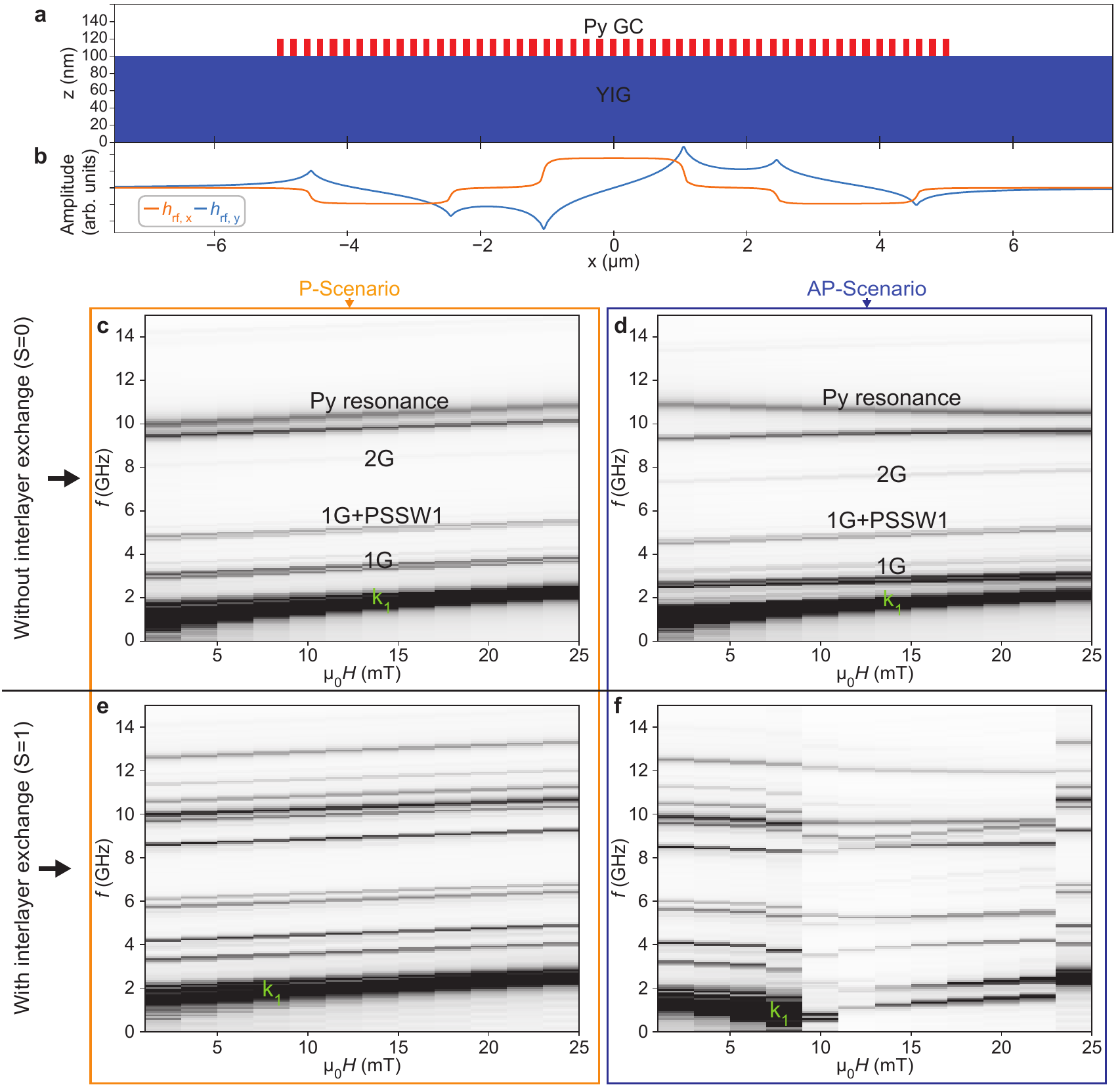}%
	\centering
	\caption{\textbf{$\mid$~Micromagnetic simulations of GC modes in P- and AP- scenario with and without interlayer coupling.}
We used Mumax3.10 \cite{doi:10.1063/1.4899186} to simulate the microwave response of the investigated GC sample assuming different interlayer exchange coupling strengths. \textbf{a} The simulated geometry consisted of a 100~nm thick YIG film (shown in blue) with a total width of $41~\upmu\mr{m}$. On top of the YIG film was a periodic grating ($a=200~\mr{nm}$) consisting of 51 Py stripes (shown in red) with a thickness of 20~nm and width of 100~nm. We simulated a slice of the GC sample in the $x-z$ plane. In $y$-direction periodic boundary condition was used (with 2048 repetitions in both $+y-$ and $-y-$direction). The grid size was $5~\mr{nm}\times5~\mr{nm}\times5~\mr{nm}$. As material parameters we used $\upmu_0 M_\mathrm{S}= 176$~mT (1005~mT), $A_\mathrm{ex}=3.75~\mathrm{pJ/m}$ ($13~\mathrm{pJ/m}$) and $\alpha=0.0009$ (0.01) for the YIG film (Py grating). \textbf{b} For SW excitation we applied a small $\bf{h_\mr{rf}}$. The spatial profile of $\bf{h_\mr{rf}}$ was simulated in COMSOL Multiphysics for the nominal CPW dimensions. Following Ref. \cite{Kumar_2017} $\bf{h_\mr{rf}}$ was modulated in time by multiplication with a Sinc-function; the SW spectra were subsequently computed from the Fourier amplitudes of the dynamic magnetization components. We simulated the SW spectra for the \textbf{(c, e)} P- and \textbf{(d, f)}  AP-scenario and for the interlayer exchange between YIG and Py either \textbf{(c, d)} completely turned off (by setting the scaling parameter $S$ in Mumax3 to zero) or \textbf{(e, f)} set to its standard value $S=1$ for which Mumax3 assumes the harmonic mean of $A_\mr{ex}$ between both layers \cite{doi:10.1063/1.4899186}. In the latter case \textbf{(e, f)}, the simulations show a strong modification of the $k_1$ mode in the AP-scenario, which was not observed in the measured spectra. We thus assume that in the experimentally investigated sample the interlayer exchange coupling was only small or completely absent. When the interlayer exchange was turned off in the simulation, we found a good qualitative agreement between simulation and measured data. In particular, the $k_1$ mode was not significantly shifted between P- and AP-scenario, while the GC modes (see labels in \textbf{(c, d)}) were red-shifted by several hundreds of MHz for the AP-scenario. We attribute this frequency shift in the absence of interlayer-exchange coupling to a modified dynamic demagnetization field. For the simulated GC modes we observed that in the AP-configuration the dynamic out-of-plane magnetization component $m_\mr{z}$ of the Py stripes was in-phase with $m_\mr{z}$ of the underlying YIG film, while in the P-configuration there was an $\pi$ phase shift between them, leading to an increased dynamic demagnetization field.
 }\label{Simul}
\end{figure*}

\begin{figure*}[bth!]
	\includegraphics[width=1\textwidth]{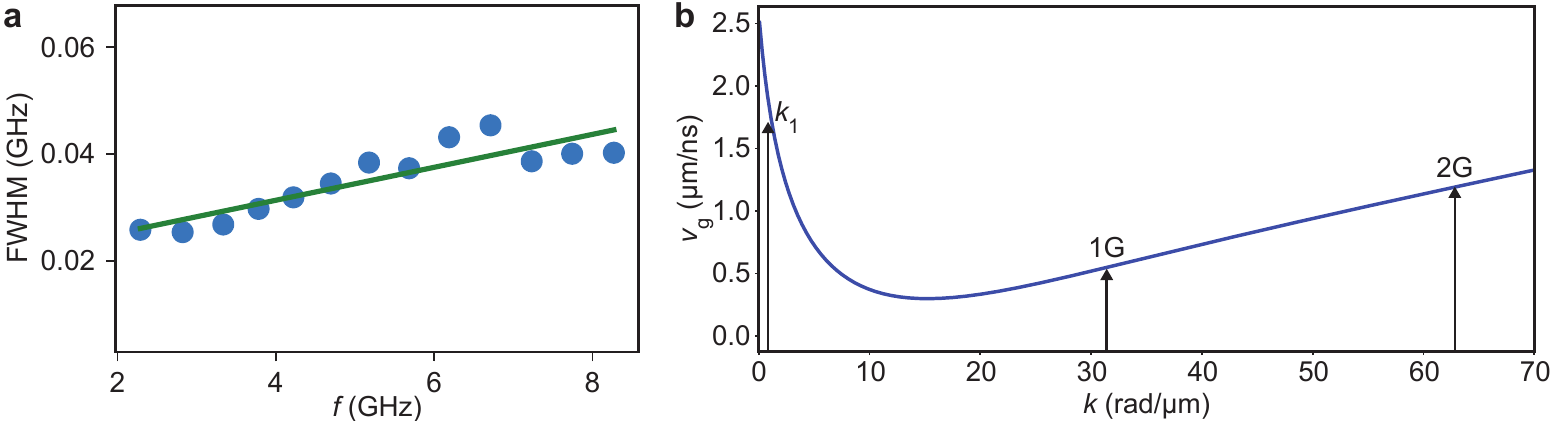}%
	\centering
	\caption{\textbf{$\mid$~Estimation of the SW decay length $l_\mr{d}$ of the $k_1$ mode at $\mu_{0}H_\mathrm{B}=14~\mr{mT}$.} \textbf{a} To extract the intrinsic Gilbert damping $\alpha_\mathrm{i}$ of the YIG film, its ferromagnetic resonance (FMR) linewidth was measured with perpendicular-to-plane magnetic fields up to 0.5~T. A CPW with a broad signal line of 20~$\upmu$m width was used, which excited spin-precessional motion at $k'_1 \simeq 0~\mr{rad}\,\upmu\mr{m}^{-1}$. The extracted full width half maximum (FWHM) of the FMR mode in Mag(S11) as a function of its resonance frequency is fitted by a linear function with a slope of $m=0.0031$ (green line) in \textbf{a}. Following Ref. \cite{doi:10.1063/1.4731273} the linewidth broadening depended on $\alpha_\mathrm{i}$ according to $\delta f = \frac{|\gamma|}{2\pi}\upmu_0\Delta H +2\alpha_\mathrm{i} f$ where $\Delta H$ was the inhomogeneous line broadening. $\delta f$ was the linewidth of the imaginary part, which was estimated by dividing the linewidth of the magnitude  Mag(S11) by a factor of $\sqrt{3}$ \cite{PhysRevLett.112.197201}. Accordingly we calculated $\alpha_\mathrm{i} = m/(2\sqrt{3})=9\times 10^{-4}$. \textbf{b} Group velocity $v_\mathrm{g}$ (line) of SWs in a 100~nm thick YIG film as calculated by the Kalinikos-Slavin formalism \cite{Kalinikos_1986}. $v_\mathrm{g}$ was calculated for $\mu_{0}H=14~\mr{mT}$ in the Damon-Eshbach configuration ($\bf{k}\perp\bf{M}_{\rm YIG}$) using a saturation magnetization $\upmu_0 M_\mathrm{YIG,S}= 176$~mT, an exchange constant $A_\mathrm{ex}=3.7~\mathrm{pJ/m}$ and $\gamma = 28.0$~GHz/T as material parameters for the YIG film. At $k_1 = 0.85~\upmu\mathrm{m/ rad}$ (marked by the left black arrow), we extract $v_\mr{g}=1.9~\upmu$m/ns. The decay length $l_\mathrm{d}$ for the SW intensity is calculated by $l_\mathrm{d}=\tau\cdot v_\mathrm{g}/2$, where $\tau=\frac{1}{2\pi\alpha f}$ is the relaxation time. Using $f=1.7$~GHz and $v_\mr{g}$ extracted for the $k_1$ mode, we find $l_\mathrm{d}=99~\upmu$m. The wave vector at 1G (2G) reads $k_{1G}=2\pi/a=2\pi/(0.2~\mu$m)~$=31.4~{\rm rad}~\mu$m$^{-1}$ ($62.8~{\rm rad}~\mu$m$^{-1}$)
	}\label{LineWith}
\end{figure*}



\begin{thebibliography}{10}
\expandafter\ifx\csname url\endcsname\relax
  \def\url#1{\texttt{#1}}\fi
\expandafter\ifx\csname urlprefix\endcsname\relax\def\urlprefix{URL }\fi
\providecommand{\bibinfo}[2]{#2}
\providecommand{\eprint}[2][]{\url{#2}}

\bibitem{Manipat2018}
\bibinfo{author}{Manipatruni, S.}, \bibinfo{author}{Nikonov, D.} \&
  \bibinfo{author}{Young, I.}
\newblock \bibinfo{title}{Beyond {CMOS} computing with spin and polarization}.
\newblock \emph{\bibinfo{journal}{Nat. Phys.}} \textbf{\bibinfo{volume}{14}},
  \bibinfo{pages}{338--343} (\bibinfo{year}{2018}).

\bibitem{Sebastian2020}
\bibinfo{author}{Sebastian, A.}, \bibinfo{author}{Gallo, M.~L.},
  \bibinfo{author}{Khaddam-Aljameh, R.} \& \bibinfo{author}{Eleftheriou, E.}
\newblock \bibinfo{title}{Memory devices and applications for in-memory
  computing}.
\newblock \emph{\bibinfo{journal}{Nat. Nanotechn.}}
  \textbf{\bibinfo{volume}{15}}, \bibinfo{pages}{529--544}
  (\bibinfo{year}{2020}).

\bibitem{BoseEin2006}
\bibinfo{author}{Demokritov, S.} \emph{et~al.}
\newblock \bibinfo{title}{Bose–{E}instein condensation of quasi-equilibrium
  magnons at room temperature under pumping}.
\newblock \emph{\bibinfo{journal}{Nature}} \textbf{\bibinfo{volume}{443}},
  \bibinfo{pages}{430} (\bibinfo{year}{2006}).

\bibitem{Khitun_2010}
\bibinfo{author}{Khitun, A.}, \bibinfo{author}{Bao, M.} \&
  \bibinfo{author}{Wang, K.~L.}
\newblock \bibinfo{title}{Magnonic logic circuits}.
\newblock \emph{\bibinfo{journal}{J. Phys. D: Appl. Phys.}}
  \textbf{\bibinfo{volume}{43}}, \bibinfo{pages}{264005}
  (\bibinfo{year}{2010}).
\newblock
  \urlprefix\url{https://doi.org/10.1088%2F0022-3727%2F43%2F26%2F264005}.

\bibitem{ChumakMagTrans}
\bibinfo{author}{Chumak, A.}, \bibinfo{author}{Serga, A.} \&
  \bibinfo{author}{Hillebrands, B.}
\newblock \bibinfo{title}{Magnon transistor for all-magnon data processing}.
\newblock \emph{\bibinfo{journal}{Nat. Commun.}} \textbf{\bibinfo{volume}{5}},
  \bibinfo{pages}{4700} (\bibinfo{year}{2014}).

\bibitem{Mahmoud2020}
\bibinfo{author}{Mahmoud, A.} \emph{et~al.}
\newblock \bibinfo{title}{Introduction to spin wave computing}.
\newblock \emph{\bibinfo{journal}{J. Appl. Phys.}}
  \textbf{\bibinfo{volume}{128}}, \bibinfo{pages}{161101}
  (\bibinfo{year}{2020}).
\newblock \urlprefix\url{https://doi.org/10.1063/5.0019328}.
\newblock \eprint{https://doi.org/10.1063/5.0019328}.

\bibitem{IRDS2020}
\bibinfo{title}{International roadmap for devices and systems ({IRDS})}.
\newblock \bibinfo{type}{Tech. Rep.} (\bibinfo{year}{2020}).
\newblock \eprint{https://irds.ieee.org/editions/2020}.

\bibitem{Islam_2019}
\bibinfo{author}{Islam, R.} \emph{et~al.}
\newblock \bibinfo{title}{Device and materials requirements for neuromorphic
  computing}.
\newblock \emph{\bibinfo{journal}{J. Phys. D: Appl. Phys.}}
  \textbf{\bibinfo{volume}{52}}, \bibinfo{pages}{113001}
  (\bibinfo{year}{2019}).
\newblock \urlprefix\url{https://doi.org/10.1088%2F1361-6463%2Faaf784}.

\bibitem{Wang2019b}
\bibinfo{author}{Wang, Y.} \emph{et~al.}
\newblock \bibinfo{title}{Magnetization switching by magnon-mediated spin
  torque through an antiferromagnetic insulator}.
\newblock \emph{\bibinfo{journal}{Science}} \textbf{\bibinfo{volume}{366}},
  \bibinfo{pages}{1125--1128} (\bibinfo{year}{2019}).

\bibitem{Han1121}
\bibinfo{author}{Han, J.}, \bibinfo{author}{Zhang, P.}, \bibinfo{author}{Hou,
  J.~T.}, \bibinfo{author}{Siddiqui, S.~A.} \& \bibinfo{author}{Liu, L.}
\newblock \bibinfo{title}{Mutual control of coherent spin waves and magnetic
  domain walls in a magnonic device}.
\newblock \emph{\bibinfo{journal}{Science}} \textbf{\bibinfo{volume}{366}},
  \bibinfo{pages}{1121--1125} (\bibinfo{year}{2019}).
\newblock \urlprefix\url{https://science.sciencemag.org/content/366/6469/1121}.
\newblock
  \eprint{https://science.sciencemag.org/content/366/6469/1121.full.pdf}.

\bibitem{Yu2014SR}
\bibinfo{author}{Yu, H.} \emph{et~al.}
\newblock \bibinfo{title}{Magnetic thin-film insulator with ultra-low spin wave
  damping for coherent nanomagnonics}.
\newblock \emph{\bibinfo{journal}{Sci. Rep.}} \textbf{\bibinfo{volume}{4}},
  \bibinfo{pages}{6848} (\bibinfo{year}{2014}).

\bibitem{Maendl2017}
\bibinfo{author}{Maendl, S.} \& \bibinfo{author}{Grundler, D.}
\newblock \bibinfo{title}{Spin waves with large decay length and few 100 nm
  wavelengths in thin yttrium iron garnet grown at the wafer scale}.
\newblock \emph{\bibinfo{journal}{Appl. Phys. Lett.}}
  \textbf{\bibinfo{volume}{111}}, \bibinfo{pages}{012403}
  (\bibinfo{year}{2017}).
\newblock \urlprefix\url{https://doi.org/10.1063/1.4991520}.
\newblock \eprint{https://doi.org/10.1063/1.4991520}.

\bibitem{Yu2016}
\bibinfo{author}{Yu, H.} \emph{et~al.}
\newblock \bibinfo{title}{Approaching soft {X}-ray wavelengths in
  nanomagnet-based microwave technology}.
\newblock \emph{\bibinfo{journal}{Nat. Commun.}} \textbf{\bibinfo{volume}{7}},
  \bibinfo{pages}{11255} (\bibinfo{year}{2016}).

\bibitem{Liu2018}
\bibinfo{author}{Liu, C.} \emph{et~al.}
\newblock \bibinfo{title}{Long-distance propagation of short-wavelength spin
  waves}.
\newblock \emph{\bibinfo{journal}{Nat. Commun.}} \textbf{\bibinfo{volume}{9}},
  \bibinfo{pages}{738} (\bibinfo{year}{2018}).

\bibitem{PhysRevB.100.104427}
\bibinfo{author}{Chen, J.} \emph{et~al.}
\newblock \bibinfo{title}{Excitation of unidirectional exchange spin waves by a
  nanoscale magnetic grating}.
\newblock \emph{\bibinfo{journal}{Phys. Rev. B}}
  \textbf{\bibinfo{volume}{100}}, \bibinfo{pages}{104427}
  (\bibinfo{year}{2019}).
\newblock \urlprefix\url{https://link.aps.org/doi/10.1103/PhysRevB.100.104427}.

\bibitem{KBaumgaertl2020}
\bibinfo{author}{Baumgaertl, K.} \emph{et~al.}
\newblock \bibinfo{title}{Nanoimaging of ultrashort magnon emission by
  ferromagnetic grating couplers at {GH}z frequencies}.
\newblock \emph{\bibinfo{journal}{Nano Lett.}} \textbf{\bibinfo{volume}{20}},
  \bibinfo{pages}{7281--7286} (\bibinfo{year}{2020}).
\newblock \urlprefix\url{https://doi.org/10.1021/acs.nanolett.0c02645}.
\newblock \bibinfo{note}{PMID: 32830984},
  \eprint{https://doi.org/10.1021/acs.nanolett.0c02645}.

\bibitem{Slon2010}
\bibinfo{author}{Slonczewski, J.~C.}
\newblock \bibinfo{title}{Initiation of spin-transfer torque by thermal
  transport from magnons}.
\newblock \emph{\bibinfo{journal}{Phys. Rev. B}} \textbf{\bibinfo{volume}{82}},
  \bibinfo{pages}{054403} (\bibinfo{year}{2010}).
\newblock \urlprefix\url{https://link.aps.org/doi/10.1103/PhysRevB.82.054403}.

\bibitem{Suresh2021}
\bibinfo{author}{Suresh, A.}, \bibinfo{author}{Bajpai, U.},
  \bibinfo{author}{Petrovi\ifmmode~\acute{c}\else \'{c}\fi{}, M.~D.},
  \bibinfo{author}{Yang, H.} \& \bibinfo{author}{Nikoli\ifmmode~\acute{c}\else
  \'{c}\fi{}, B.~K.}
\newblock \bibinfo{title}{Magnon- versus electron-mediated spin-transfer torque
  exerted by spin current across an antiferromagnetic insulator to switch the
  magnetization of an adjacent ferromagnetic metal}.
\newblock \emph{\bibinfo{journal}{Phys. Rev. Applied}}
  \textbf{\bibinfo{volume}{15}}, \bibinfo{pages}{034089}
  (\bibinfo{year}{2021}).
\newblock
  \urlprefix\url{https://link.aps.org/doi/10.1103/PhysRevApplied.15.034089}.

\bibitem{Guo2021}
\bibinfo{author}{Guo, C.~Y.} \emph{et~al.}
\newblock \bibinfo{title}{Switching the perpendicular magnetization of a
  magnetic insulator by magnon transfer torque}.
\newblock \emph{\bibinfo{journal}{Phys. Rev. B}}
  \textbf{\bibinfo{volume}{104}}, \bibinfo{pages}{094412}
  (\bibinfo{year}{2021}).
\newblock \urlprefix\url{https://link.aps.org/doi/10.1103/PhysRevB.104.094412}.

\bibitem{Zheng2022}
\bibinfo{author}{Zheng, D.} \emph{et~al.}
\newblock \bibinfo{title}{High-efficiency magnon-mediated magnetization
  switching in all-oxide heterostructures with perpendicular magnetic
  anisotropy}.
\newblock \emph{\bibinfo{journal}{Advanced Materials}}
  \textbf{\bibinfo{volume}{n/a}}, \bibinfo{pages}{2203038}.
\newblock
  \urlprefix\url{https://onlinelibrary.wiley.com/doi/abs/10.1002/adma.202203038}.
\newblock
  \eprint{https://onlinelibrary.wiley.com/doi/pdf/10.1002/adma.202203038}.

\bibitem{Papp2021}
\bibinfo{author}{Papp, {\'A}.}, \bibinfo{author}{Porod, W.} \&
  \bibinfo{author}{Csaba, G.}
\newblock \bibinfo{title}{Nanoscale neural network using non-linear spin-wave
  interference}.
\newblock \emph{\bibinfo{journal}{Nat. Commun.}} \textbf{\bibinfo{volume}{12}},
  \bibinfo{pages}{6422} (\bibinfo{year}{2021}).

\bibitem{Qin2021}
\bibinfo{author}{Qin, H.}, \bibinfo{author}{Holländer, R.} \&
  \bibinfo{author}{Flajšman, {\em et al.}., L.}
\newblock \bibinfo{title}{Nanoscale magnonic {F}abry-{P}érot resonator for
  low-loss spin-wave manipulation.}
\newblock \emph{\bibinfo{journal}{Nat. Commun.}} \textbf{\bibinfo{volume}{12}},
  \bibinfo{pages}{2293} (\bibinfo{year}{2021}).

\bibitem{Yu2013}
\bibinfo{author}{Yu, H.} \emph{et~al.}
\newblock \bibinfo{title}{Omnidirectional spin-wave nanograting coupler}.
\newblock \emph{\bibinfo{journal}{Nat. Commun.}} \textbf{\bibinfo{volume}{4}},
  \bibinfo{pages}{2702} (\bibinfo{year}{2013}).

\bibitem{Kostylev2005}
\bibinfo{author}{Kostylev, M.~P.}, \bibinfo{author}{Serga, A.~A.},
  \bibinfo{author}{Schneider, T.}, \bibinfo{author}{Leven, B.} \&
  \bibinfo{author}{Hillebrands, B.}
\newblock \bibinfo{title}{Spin-wave logical gates}.
\newblock \emph{\bibinfo{journal}{Appl. Phys. Lett.}}
  \textbf{\bibinfo{volume}{87}}, \bibinfo{pages}{153501}
  (\bibinfo{year}{2005}).

\bibitem{Schneider2008}
\bibinfo{author}{Schneider, T.} \emph{et~al.}
\newblock \bibinfo{title}{Realization of spin-wave logic gates}.
\newblock \emph{\bibinfo{journal}{Appl. Phys. Lett.}}
  \textbf{\bibinfo{volume}{92}}, \bibinfo{pages}{022505}
  (\bibinfo{year}{2008}).

\bibitem{Csaba2017}
\bibinfo{author}{Csaba, G.}, \bibinfo{author}{Papp, {\'{A}}.} \&
  \bibinfo{author}{Porod, W.}
\newblock \bibinfo{title}{Perspectives of using spin waves for computing and
  signal processing}.
\newblock \emph{\bibinfo{journal}{Phys. Lett. A}}
  \textbf{\bibinfo{volume}{381}}, \bibinfo{pages}{1471--1476}
  (\bibinfo{year}{2017}).

\bibitem{Maendl2018}
\bibinfo{author}{Maendl, S.} \& \bibinfo{author}{Grundler, D.}
\newblock \bibinfo{title}{Multi-directional emission and detection of spin
  waves propagating in yttrium iron garnet with wavelengths down to about 100
  nm}.
\newblock \emph{\bibinfo{journal}{Appl. Phys. Lett.}}
  \textbf{\bibinfo{volume}{112}}, \bibinfo{pages}{192410}
  (\bibinfo{year}{2018}).
\newblock \urlprefix\url{https://doi.org/10.1063/1.5026060}.
\newblock \eprint{https://doi.org/10.1063/1.5026060}.

\bibitem{Nozaki2007}
\bibinfo{author}{Nozaki, Y.} \emph{et~al.}
\newblock \bibinfo{title}{Microwave-assisted magnetization reversal in
  0.36-$\upmu$m-wide permalloy wires}.
\newblock \emph{\bibinfo{journal}{Appl. Phys. Lett.}}
  \textbf{\bibinfo{volume}{91}}, \bibinfo{pages}{122505}
  (\bibinfo{year}{2007}).

\bibitem{Topp2009}
\bibinfo{author}{Topp, J.}, \bibinfo{author}{Heitmann, D.} \&
  \bibinfo{author}{Grundler, D.}
\newblock \bibinfo{title}{Interaction effects on microwave-assisted switching
  of {N}i$_{80}${F}e$_{20}$ nanowires in densely packed arrays}.
\newblock \emph{\bibinfo{journal}{Phys. Rev. B}} \textbf{\bibinfo{volume}{80}},
  \bibinfo{pages}{174421} (\bibinfo{year}{2009}).

\bibitem{6466529}
\bibinfo{author}{{Fallarino}, L.} \emph{et~al.}
\newblock \bibinfo{title}{Propagation of spin waves excited in a permalloy film
  by a finite-ground coplanar waveguide: A combined phase-sensitive
  micro-focused {B}rillouin light scattering and micromagnetic study}.
\newblock \emph{\bibinfo{journal}{IEEE Trans. Magn.}}
  \textbf{\bibinfo{volume}{49}}, \bibinfo{pages}{1033--1036}
  (\bibinfo{year}{2013}).

\bibitem{PhysRevB.86.054414}
\bibinfo{author}{Schultheiss, H.}, \bibinfo{author}{Vogt, K.} \&
  \bibinfo{author}{Hillebrands, B.}
\newblock \bibinfo{title}{Direct observation of nonlinear four-magnon
  scattering in spin-wave microconduits}.
\newblock \emph{\bibinfo{journal}{Phys. Rev. B}} \textbf{\bibinfo{volume}{86}},
  \bibinfo{pages}{054414} (\bibinfo{year}{2012}).
\newblock \urlprefix\url{https://link.aps.org/doi/10.1103/PhysRevB.86.054414}.

\bibitem{doi:10.1063/1.2756481}
\bibinfo{author}{Olson, H.~M.}, \bibinfo{author}{Krivosik, P.},
  \bibinfo{author}{Srinivasan, K.} \& \bibinfo{author}{Patton, C.~E.}
\newblock \bibinfo{title}{Ferromagnetic resonance saturation and second order
  suhl spin wave instability processes in thin permalloy films}.
\newblock \emph{\bibinfo{journal}{J. Appl. Phys.}}
  \textbf{\bibinfo{volume}{102}}, \bibinfo{pages}{023904}
  (\bibinfo{year}{2007}).
\newblock \urlprefix\url{https://doi.org/10.1063/1.2756481}.
\newblock \eprint{https://doi.org/10.1063/1.2756481}.

\bibitem{PhysRevLett.99.207202}
\bibinfo{author}{Podbielski, J.}, \bibinfo{author}{Heitmann, D.} \&
  \bibinfo{author}{Grundler, D.}
\newblock \bibinfo{title}{Microwave-assisted switching of microscopic rings:
  Correlation between nonlinear spin dynamics and critical microwave fields}.
\newblock \emph{\bibinfo{journal}{Phys. Rev. Lett.}}
  \textbf{\bibinfo{volume}{99}}, \bibinfo{pages}{207202}
  (\bibinfo{year}{2007}).
\newblock
  \urlprefix\url{https://link.aps.org/doi/10.1103/PhysRevLett.99.207202}.

\bibitem{Thirion2003}
\bibinfo{author}{Thirion, C.}, \bibinfo{author}{Wernsdorfer, W.} \&
  \bibinfo{author}{Mailly, D.}
\newblock \bibinfo{title}{Switching of magnetization by nonlinear resonance
  studied in single nanoparticles}.
\newblock \emph{\bibinfo{journal}{Nat. Mater.}} \textbf{\bibinfo{volume}{2}},
  \bibinfo{pages}{524} (\bibinfo{year}{2003}).

\bibitem{Nembach2007}
\bibinfo{author}{Nembach, H.~T.} \emph{et~al.}
\newblock \bibinfo{title}{Microwave assisted switching in a
  {N}i$_{81}${F}e$_{19}$ ellipsoid}.
\newblock \emph{\bibinfo{journal}{Appl. Phys. Lett.}}
  \textbf{\bibinfo{volume}{90}}, \bibinfo{pages}{062503}
  (\bibinfo{year}{2007}).
\newblock \urlprefix\url{https://doi.org/10.1063/1.2450645}.
\newblock \eprint{https://doi.org/10.1063/1.2450645}.

\bibitem{Schoen2016}
\bibinfo{author}{Schoen, M. A.~W.} \emph{et~al.}
\newblock \bibinfo{title}{Ultra-low magnetic damping of a
  metallic~ferromagnet}.
\newblock \emph{\bibinfo{journal}{Nat. Phys.}} \textbf{\bibinfo{volume}{12}},
  \bibinfo{pages}{839--842} (\bibinfo{year}{2016}).

\bibitem{Hyde2014}
\bibinfo{author}{Hyde, P.} \emph{et~al.}
\newblock \bibinfo{title}{Electrical detection of direct and alternating spin
  current injected from a ferromagnetic insulator into a ferromagnetic metal}.
\newblock \emph{\bibinfo{journal}{Phys. Rev. B}} \textbf{\bibinfo{volume}{89}},
  \bibinfo{pages}{180404} (\bibinfo{year}{2014}).
\newblock \urlprefix\url{https://link.aps.org/doi/10.1103/PhysRevB.89.180404}.

\bibitem{Iguchi2013}
\bibinfo{author}{Iguchi, R.} \emph{et~al.}
\newblock \bibinfo{title}{Spin pumping by nonreciprocal spin waves under local
  excitation}.
\newblock \emph{\bibinfo{journal}{Appl. Phys. Lett.}}
  \textbf{\bibinfo{volume}{102}}, \bibinfo{pages}{022406}
  (\bibinfo{year}{2013}).
\newblock \urlprefix\url{https://doi.org/10.1063/1.4775685}.
\newblock \eprint{https://doi.org/10.1063/1.4775685}.

\bibitem{SandwegSaitoh2010}
\bibinfo{author}{Sandweg, C.~W.}, \bibinfo{author}{Kajiwara, Y.},
  \bibinfo{author}{Ando, K.}, \bibinfo{author}{Saitoh, E.} \&
  \bibinfo{author}{Hillebrands, B.}
\newblock \bibinfo{title}{Enhancement of the spin pumping efficiency by spin
  wave mode selection}.
\newblock \emph{\bibinfo{journal}{Appl. Phys. Lett.}}
  \textbf{\bibinfo{volume}{97}}, \bibinfo{pages}{252504}
  (\bibinfo{year}{2010}).
\newblock \urlprefix\url{https://doi.org/10.1063/1.3528207}.
\newblock \eprint{https://doi.org/10.1063/1.3528207}.

\bibitem{Mucchietto2022}
\bibinfo{author}{Mucchietto, A.}, \bibinfo{author}{Baumgaertl, K.} \&
  \bibinfo{author}{Grundler, D.}
\newblock \bibinfo{title}{unpublished}.

\bibitem{Kruglyak2021}
\bibinfo{author}{Kruglyak, V.~V.}
\newblock \bibinfo{title}{Chiral magnonic resonators: Rediscovering the basic
  magnetic chirality in magnonics}.
\newblock \emph{\bibinfo{journal}{Appl. Phys. Lett.}}
  \textbf{\bibinfo{volume}{119}}, \bibinfo{pages}{200502}
  (\bibinfo{year}{2021}).
\newblock \urlprefix\url{https://doi.org/10.1063/5.0068820}.
\newblock \eprint{https://doi.org/10.1063/5.0068820}.

\bibitem{Albert2000}
\bibinfo{author}{Albert, F.~J.}, \bibinfo{author}{Katine, J.~A.},
  \bibinfo{author}{Buhrman, R.~A.} \& \bibinfo{author}{Ralph, D.~C.}
\newblock \bibinfo{title}{Spin-polarized current switching of a {C}o thin film
  nanomagnet}.
\newblock \emph{\bibinfo{journal}{Applied Physics Letters}}
  \textbf{\bibinfo{volume}{77}}, \bibinfo{pages}{3809--3811}
  (\bibinfo{year}{2000}).
\newblock \urlprefix\url{https://doi.org/10.1063/1.1330562}.
\newblock \eprint{https://doi.org/10.1063/1.1330562}.

\bibitem{doi:10.1063/1.4899186}
\bibinfo{author}{Vansteenkiste, A.} \emph{et~al.}
\newblock \bibinfo{title}{The design and verification of {MuMax3}}.
\newblock \emph{\bibinfo{journal}{AIP Advances}} \textbf{\bibinfo{volume}{4}},
  \bibinfo{pages}{107133} (\bibinfo{year}{2014}).
\newblock \eprint{https://doi.org/10.1063/1.4899186}.

\bibitem{Kumar_2017}
\bibinfo{author}{Kumar, D.} \& \bibinfo{author}{Adeyeye, A.~O.}
\newblock \bibinfo{title}{Techniques in micromagnetic simulation and analysis}.
\newblock \emph{\bibinfo{journal}{J. Phys. D: Appl. Phys.}}
  \textbf{\bibinfo{volume}{50}}, \bibinfo{pages}{343001}
  (\bibinfo{year}{2017}).

\bibitem{doi:10.1063/1.4731273}
\bibinfo{author}{Yu, H.} \emph{et~al.}
\newblock \bibinfo{title}{High propagating velocity of spin waves and
  temperature dependent damping in a {CoFeB} thin film}.
\newblock \emph{\bibinfo{journal}{Appl. Phys. Lett.}}
  \textbf{\bibinfo{volume}{100}}, \bibinfo{pages}{262412}
  (\bibinfo{year}{2012}).
\newblock \eprint{https://doi.org/10.1063/1.4731273}.

\bibitem{PhysRevLett.112.197201}
\bibinfo{author}{Wang, H.~L.} \emph{et~al.}
\newblock \bibinfo{title}{Scaling of {S}pin {H}all {A}ngle in 3d, 4d, and 5d
  {M}etals from
  {${\mathrm{{Y}}}_{3}{\mathrm{{F}e}}_{5}{\mathrm{{O}}}_{12}$}/{M}etal {S}pin
  {P}umping}.
\newblock \emph{\bibinfo{journal}{Phys. Rev. Lett.}}
  \textbf{\bibinfo{volume}{112}}, \bibinfo{pages}{197201}
  (\bibinfo{year}{2014}).

\bibitem{Kalinikos_1986}
\bibinfo{author}{Kalinikos, B.~A.} \& \bibinfo{author}{Slavin, A.~N.}
\newblock \bibinfo{title}{Theory of dipole-exchange spin wave spectrum for
  ferromagnetic films with mixed exchange boundary conditions}.
\newblock \emph{\bibinfo{journal}{J. Phys. C: Solid St. Phys.}}
  \textbf{\bibinfo{volume}{19}}, \bibinfo{pages}{7013--7033}
  (\bibinfo{year}{1986}).

\end{thebibliography}
\end{document}